# Differentiation of Planetesimals and the Thermal Consequences of Melt Migration

Nicholas Moskovitz

*Department of Terrestrial Magnetism, Carnegie Institution of Washington*
*5241 Broad Branch Road, Washington, DC 20015*

`nmoskovitz@dtm.ciw.edu`

Eric Gaidos

*Department of Geology and Geophysics, University of Hawaii*
*1680 East-West Road, Honolulu, HI 96822*

**ABSTRACT**

We model the heating of a primordial planetesimal by decay of the short-lived radionuclides $^{26}$Al and $^{60}$Fe to determine (i) the timescale on which melting will occur; (ii) the minimum size of a body that will produce silicate melt and differentiate; (iii) the migration rate of molten material within the interior; and (iv) the thermal consequences of the transport of $^{26}$Al in partial melt. Our models incorporate results from previous studies of planetary differentiation and are constrained by petrologic (i.e. grain size distributions), isotopic (e.g. $^{207}$Pb-$^{206}$Pb and $^{182}$Hf-$^{182}$W ages) and mineralogical properties of differentiated achondrites. We show that formation of a basaltic crust via melt percolation was limited by the formation time of the body, matrix grain size and viscosity of the melt. We show that low viscosity (< 1 Pa·s) silicate melt can buoyantly migrate on a timescale comparable to the mean life of $^{26}$Al. The equilibrium partitioning of Al into silicate partial melt and the migration of that melt acts to dampen internal temperatures. However, subsequent heating from the decay of $^{60}$Fe generated melt fractions in excess of 50%, thus completing differentiation for bodies that accreted within 2 Myr of CAI formation (i.e. the onset of isotopic decay). Migration and concentration of $^{26}$Al into a crust results in remelting of that crust for accretion times less than 2 Myr and for bodies >100 km in size. Differentiation would be most likely for planetesimals larger than 20 km in diameter that accreted within ~2.7 Myr of CAI formation.

## 1. Introduction

Planetesimals that accreted within a few Myr after the condensation of the first Solar System solids (calcium-aluminum-rich inclusions, CAIs) are presumed to have been homogenous bodies, with chemical compositions similar to chondrites

(Lodders 2003). The interiors of these objects would have been heated by the decay of short-lived radioactive isotopes (SLRs, Urey 1955). Other heat sources, such as the energy of accretion and impacts, were insignificant relative to SLR decay for bodies <1000 km in diameter (Moskovitz 2009). Electromagnetic induction heating has been proposed as a heat source for bodies with metallic iron cores (Sonett et al. 1968), but this mechanism is poorly constrained (Wood et al. 1991) and is inconsistent with the geometry of T-Tauri outflows (McSween et al. 2002), thus it will not be considered here.

While multiple SLRs may have been present in the early Solar System (Huss et al. 2009), only two, $^{26}$Al and $^{60}$Fe, were potentially significant sources of internal heat. Most, but not all CAIs contained $^{26}$Al at a concentration close to the canonical ratio $^{26}$Al/$^{27}$Al ~ 4-5×10$^{-5}$ (Jacobsen et al. 2008; MacPherson et al. 2010). The consistent enhancement in $^{26}$Mg of planetary materials with respect to CAIs suggests that $^{26}$Al was uniformly distributed in the solar system (Thrane et al. 2006; Villeneuve et al. 2009), although some refractory grains lack $^{26}$Al and there may have been $^{26}$Al-free reservoirs. Near-homogenous $^{26}$Al supports an exogenous origin such as a supernova or Wolf-Rayet star (Ouellette et al. 2005; Gaidos et al. 2009), as opposed to the active young Sun (Gounelle et al. 2006). The initial abundance of $^{60}$Fe relative to $^{56}$Fe is more difficult to infer and a reliable value has proven elusive, but was perhaps no more than ~10$^{-6}$ (Tachbana et al. 2006; Dauphas et al. 2008; Mishra et al. 2010). This isotope also seems to have been uniformly distributed throughout the Solar System (Dauphas et al. 2008), and the initial value may be typical of star-forming regions (Gounelle et al. 2009). Table 1 gives the adopted initial concentration and half-life of each SLR.

These concentrations correspond to an integrated energy input of 6.7 x 10$^6$ J/kg for $^{26}$Al and 7.0 x 10$^5$ J/kg for $^{60}$Fe, assuming chondritic elemental abundances (0.0113 and 0.24 respectively; Ghosh et al. 1998). These energies were sufficient to melt the interior of planetesimals: the Fe-S eutectic at 1200 K marks the onset of melting (McCoy et al. 2006); 1400 and 1900 K are typical solidus and liquidus temperatures for planetary silicates (McKenzie et al. 1988; Agee 1997); and the Fe-Ni metal that dominates the melt compositions of iron meteorites at 1800 K (Benedix et al. 2000). Partially molten metals and silicates can gravitationally segregate, forming differentiated interiors. Subsequent collisional disruption of a differentiated planetesimal would have produced a suite of fragments, some of which are recovered today as meteorites, ranging from pieces of Fe-Ni core, to dunite-rich chunks of mantle, to feldspathic crustal basalt (Mittlefehldt et al. 1998).

Both meteorites and asteroids provide evidence for widespread differentiation in the early Solar System. The majority of the ~150 parent bodies represented in meteorite collections experienced metal-silicate differentiation (Burbine et al. 2002). The canonical example of a differentiated asteroid is the large (~500 km) asteroid 4 Vesta. The spectrum of Vesta closely matches the basaltic HED meteorites, thereby suggesting that Vesta's surface is an undisrupted differentiated crust (McCord et al. 1970). This spectral similarity, the lack of other large basaltic asteroids in the main belt, Vesta's favorable location for the delivery of fragments to Earth, and the presence of an extensive collisional family of Vestoids (Binzel et al. 1993) suggest that the HEDs are genetically related to Vesta (Consolmagno et al. 1977).

The thermal history and differentiation of Vesta-like bodies have been modeled using spherically symmetric descriptions of heat conduction. Ghosh et al. (1998) modeled Vesta's thermal evolution through core and crust formation, assuming instantaneous segregation of metal and silicate phases at the metal-sulfide and silicate liquidi respectively. These calculations constrained the epochs of accretion (2.85 Myr after CAIs), core formation (4.58 Myr), crust formation (6.58 Myr), and geochemical closure of Vesta (~100 Myr), all of which are broadly consistent with the measured chronologies of HED meteorites. In addition, Ghosh et al. (1998) showed that concentration of $^{26}$Al into Vesta's crust affected the body's thermal evolution, while the decay of $^{60}$Fe following its segregation into the core, did not. However, the abundance of $^{60}$Fe adopted by these authors was more than an order of magnitude less than current estimates (e.g. Mishra et al. 2010). Hevey et al. (2006) presented a series of models for the heating of a planetesimal by the decay of $^{26}$Al and traced these bodies' evolution through sintering of the initially unconsolidated silicates, and the formation and convection of magma oceans. They found that the parent bodies of differentiated meteorites must have accreted within ~2 Myr of CAI formation, but these authors did not take into account the consequences of SLR redistribution. Sahijpal et al. (2007) coupled a planetesimal accretion model with a thermal conduction model, and included volume loss due to sintering, redistribution of SLRs due to melting, and the gradual formation of both a core and crust. This study also found that differentiation occurred on bodies that accreted within 2-3 Myr of CAI formation. In general, these works all suggest that a time of accretion <3 Myr was necessary for differentiation. No general consensus was reached regarding the importance of $^{60}$Fe, in part due to different values used for its initial abundance.

Other work has described the dynamics of the melting and differentiation process. Taylor (1992) described core formation on asteroids based on the analytical works of McKenzie (1984, 1985) and Stevenson (1990). He concluded that high degrees of partial melting (~50%) were required for metal to efficiently segregate into the cores of iron meteorite parent bodies. Wilson et al. (1991) proposed that exsolution of <1 wt% of $H_2O$ and $CO_2$ would increase melt buoyancy to an extent that basaltic melt would pyroclastically erupt at speeds greater than the local escape velocity for bodies smaller than approximately 100 km. Objects larger than this (e.g. asteroid 4 Vesta) had sufficient surface gravity to retain erupted melt in a crust. Such eruptions could also occur following the pooling of basaltic liquid in large sills at the base of an unaltered, less permeable layer in the upper ~5 km (Wilson et al. 2008). This model explains the lack of a basaltic component associated with the aubrite, acapulcoite-lodranite and ureilite meteorites, each of which derived from parent bodies that experienced partial melting (Wilson et al. 1991; McCoy et al. 1997; Goodrich et al. 2007).

Here we address several aspects of differentiation not considered by these previous studies. These include a realistic treatment of silicate melt migration rates and the thermal consequences of such migration, constraints on the initial conditions (e.g. time of accretion) responsible for the formation of putative magma oceans on planetesimals like the asteroid 4 Vesta (Righter et al. 1997) and the parent bodies of magmatic iron meteorites (Scott 1972), and quantifying the influence of $^{60}$Fe decay on differentiation. Recently updated constraints on initial SLR abundances in the Solar System (Jacobsen et al. 2008; Tachibana et al. 2006; Mishra et al. 2010) provide additional motivation for revisiting issues related to planetesimal differentiation.

In Section 2 we describe physical processes treated by our collection of models: thermal conduction in a spherical body; the production and migration of partial melt; and the transport of the primary heat source, $^{26}$Al, in that melt. In Section 3 we present calculations for the thermal evolution of a planetesimal up to differentiation, the influence of melt migration, and an estimate of the minimum size for differentiation. In Section 4 we describe and justify the exclusion of several processes from our model. Finally, in Section 5 we combine these results in a synthesis of the relevant processes, time scales and sizes associated with differentiated bodies, compare our results with previous studies, and propose future areas of investigation.

Table 1 lists the variables and parameters employed in our calculations.

## 2. Models of Physical Processes

### 2.1 Thermal Conduction

Conduction will remove heat from the interior of a planetesimal and impose a minimum size for differentiation. We assume 1-dimensional heat conduction in a spherically symmetric body with internal heating (Grimm et al. 1989):

$$\frac{\partial T}{\partial t} = \frac{1}{r^2}\frac{\partial}{\partial r}\left(r^2 \kappa \frac{\partial T}{\partial r}\right) + \frac{Q(t)}{c}, \qquad (1)$$

where $r$ is the radial coordinate, $Q(t)$ is the decay energy per unit mass per unit time for $^{26}$Al and $^{60}$Fe, and $c$ is the specific heat at constant pressure. The thermal diffusivity is related to the thermal conductivity $k$, the density $\rho$ and specific heat $c$:

$$\kappa = \frac{k}{\rho c}. \qquad (2)$$

Assuming a radially constant $\rho$, $\kappa$ and $c$, and expanding Equation 1 yields the conduction equation for $r > 0$:

$$\frac{\partial T}{\partial t} = \frac{2k}{r\rho c}\frac{\partial T}{\partial r} + \frac{k}{\rho c}\frac{\partial^2 T}{\partial r^2} + \frac{Q(t)}{c}. \qquad (3)$$

The singularity at $r=0$ can be remedied by expanding the offending $1/r$ term in a first-order Taylor series, yielding the conduction equation for $r=0$:

$$\frac{\partial T}{\partial t} = \frac{3k}{\rho c}\frac{\partial^2 T}{\partial r^2} + \frac{Q(t)}{c}. \qquad (4)$$

Equations 3 and 4 are solved using the explicit finite difference method (Ozisik 1994) with a constant temperature prescribed for the outer boundary ($r=R$). Though a radiative boundary condition offers a more robust approach (Ghosh et al. 1998), the errors associated with a simpler fixed-temperature, Dirichlet boundary condition are small because radiation balance for bodies in the inner Solar System will always dominate interior heat flow (Hevey et al. 2006). The initial temperature of the body is assumed to be constant for all radii and equal to the boundary temperature. The spatial element size and time step are chosen sufficiently small (e.g. 250 m and 800 yr) to ensure stability of the numerical result. For each time step heating from $^{26}$Al and $^{60}$Fe decay is computed:

$$Q(t) = f_{Al}\left[\frac{^{26}Al}{^{27}Al}\right]\frac{E_{Al}}{\tau_{Al}}e^{-t/\tau_{Al}} + f_{Fe}\left[\frac{^{60}Fe}{^{56}Fe}\right]\frac{E_{Fe}}{\tau_{Fe}}e^{-t/\tau_{Fe}}, \qquad (5)$$

where $t$ is time relative to CAI formation and all other quantities are defined in Table 1.

For temperatures above the silicate solidus, the effective heat capacity $c$ is computed as an average, weighted by the melt fraction, of the specific heats of the solid and liquid phases (800 and 2000 J/kg/K respectively; Ghosh et al. 1998). These temperature-independent values are upper limits for chondritic assemblages (Ghosh et al. 1999; Navrotsky 1995), their adoption produces temperatures that are underestimated by less than ~10% (Sec. 4.5). The melt fraction $\Phi$ is calculated from an empirical fit to the fertility function of peridotite (McKenzie et al. 1988):

$$\Phi(T') = 0.3936 + 0.253 \cdot T' + 0.4256 \cdot T'^2 + 2.988 \cdot T'^3 \qquad (6)$$

and

$$T' = \frac{T - 1418.1}{636.2}. \qquad (7)$$

The coefficients in these equations were calculated from the solidus and liquidus temperatures of peridotite at a pressure of 1 bar (1373 and 2009 K respectively, McKenzie et al. 1988). Planetesimals smaller than 500 km had internal pressures less than 1 kbar (assuming a spherical body in hydrostatic equilibrium with a mean density of 3300 kg/m$^3$) and would not have dramatically different solidus and liquidus points. The latent heat of fusion $L$ for chondritic silicates is $4\times10^5$ J/kg (Ghosh et al. 1998) and was accounted for between the silicate solidus and liquidus based on the melt fraction computed from Equation 6.

*2.2 Melt Migration*

A basaltic crust can form either as a result of migration of silicate partial melts to the surface (Walker et al. 1978; McKenzie 1984, 1985; Taylor et al. 1993) or through fractional crystallization of a magma ocean (McCoy et al. 2006). This second mechanism is not considered here, because it is likely that melt migration would have preceded the formation of a magma ocean (Sec. 3.2).

Molten silicates will migrate in the presence of a density contrast between the melt and the surrounding solid matrix, if the melt is interconnected in a network of pore spaces between the solid grains, and if the matrix compacts to expel the melt (i.e. to avoid the formation of vacuum pore space). Typical density contrasts for silicate partial melts $\Delta\rho$ are 300 - 700 kg/m$^3$; we adopt the lower value to minimize the rate of melt migration. Experimental and theoretical work has shown that even at melt fractions of a few percent, molten silicates form an interconnected network (Taylor et al. 1993). Compaction of a melted region is controlled by the bulk viscosity of the matrix, microscopic shear viscosity of the matrix, viscosity of the melt and permeability of the matrix to fluid flow control (McKenzie 1984, 1985). Assuming reasonable values for these material properties, the size of the compacted region ($\delta_c$) can be parameterized in terms of melt fraction, grain size and viscosity of the melt ($\phi$, $a$ and $\mu$ respectively; McKenzie 1985):

$$\delta_c = (10km)\phi^{3/2}\left(\frac{a}{1mm}\right)\left(\frac{10Pa\cdot s}{\mu}\right)^{1/2}. \qquad (8)$$

In general, compaction-driven migration will occur if the thickness of the region undergoing compaction ($\delta_c$) is smaller than the melting region (Taylor et al. 1993). The size of the melting zone for a planetesimal uniformly heated by radionuclide decay is comparable to its radius $R$ (Wilson et al. 2008). When $R/\delta_c$ is much greater than unity, compaction is likely to occur.

The four key parameters for compaction-driven melt migration ($R$, $\mu$, $a$ and $\phi$) can each vary by several orders of magnitude, thus making it difficult to define specific conditions for compaction. However, inserting plausible values into Equation 8 demonstrates that compaction and melt migration were likely outcomes.

Chondrules, which offer the best available guide for grain sizes in unaltered planetesimals (Taylor et al. 1993), are typically several 100 μm in size and are rarely larger than 1 cm (Weisberg et al. 2006). The viscosity of molten silicates varies from 0.001 to 1000 Pa·s, however, basaltic melts derived from chondritic precursors have silica contents generally ≤ 50% (Mittlefehldt et al. 1998) and viscosities of ~1-100 Pa·s (Giordano et al. 2008). Bodies ~100 km in size with $a < 1$ cm and $\mu > 1$ Pa·s, will expel melt up to $\phi = 50\%$ (Eqn. 8). Equation 8 suggests that a basaltic crust can form via compaction for bodies as small as ~20 km in size, assuming a 10% melt fraction (i.e. consistent with the degree of partial melting that removed basalt from the lodranite meteorites; McCoy et al. 1997). This size limit is similar to that for the onset of buoyant melt migration within a rigid (non-compacting) silicate matrix (Walker et al. 1978). Fully molten bodies smaller than 20 km may develop a basaltic crust via fractional crystallization of a magma ocean.

Employing Equations 2 and 4 of Taylor et al. (1993) and Equation 7 from McKenzie (1985), the characteristic e-folding time for the expulsion of molten silicates from a compacted region is:

$$\tau_{mig} = \frac{3000\mu}{4\pi a^2 \phi^2 (1-\phi)^2 \Delta\rho G \bar{\rho}} \ , \tag{9}$$

where all parameters are the same as in Equation 8 and/or defined in Table 1. This equation is valid when the size of the melt zone is comparable to $R$, the body is of uniform density, and $R/\delta_c \gg 1$. Interestingly, $\tau_{mig}$ is independent of planetesimal size: the low velocity of melt migration in small bodies (due to low gravitational acceleration) is counter-acted by the short distance over which the melt must travel. For standard values of $\Delta\rho$ and $\bar{\rho}$ (Table 1), $\mu = 1$ Pa·s (typical of basaltic melts), and when grains grow larger than approximately 1 mm, the migration time scale becomes comparable to the $^{26}$Al mean-life (Eqn 9).

### 2.3 Removal of $^{26}$Al Via Melt Migration

Aluminum preferentially partitions out of solid phases and into the early stages of silicate partial melts (Agee et al. 1990; Miyamoto et al. 1994; Pack et al. 2003). Migration of these melts can influence internal heating due to redistribution of $^{26}$Al. Primitive and differentiated achondrites derived from partial melt residues exhibit decreasing Al abundance with increasing degrees of melting (Mittlefehldt et al. 1998), a trend consistent with the early removal of Al-enriched partial melts.

A zero-dimensional box model is developed to quantify the thermal consequences of the removal of Al-enriched partial melt. Only unmelted silicates and pore space are initially present and as temperature increases from the heat of $^{26}$Al decay, partial melt is produced and migrates out of the box on a characteristic time scale (Equation 9). We assume that the planetesimal is sufficiently large such that the rate of heat loss due to conduction is negligible relative to timescales of heating and melt migration.

The amount of melt in the box is governed by:

$$\frac{d\phi}{dt} = -\frac{\phi}{\tau_{mig}} + \frac{\partial \Phi(T)}{\partial T}\frac{\partial T}{\partial t}, \qquad (10)$$

where $\tau_{mig}$ is given by Equation 9 and all other parameters are given in Table 1. Equation 10 accounts for the melt lost due to migration and the amount produced due to heating. By defining

$$\alpha = \frac{3000\mu}{4\pi a^2 G \bar{\rho} \Delta \rho}, \qquad (11)$$

Equation 10 can be rewritten as:

$$\frac{d\phi}{dt} = -\frac{\phi^3(1-\phi)^2}{\alpha} + \frac{\partial \Phi}{\partial T}\frac{\partial T}{\partial t}. \qquad (12)$$

All of physical properties of this system are included in the α-parameter and with the exception of grain size ($a$) are assumed constant. The coarsening of grains associated with increasing degrees of partial melting (i.e. Ostwald ripening; Taylor 1992) is parameterized using a correlation between typical grain size and degree of partial melting for several achondrite meteorite groups (see Mittlefehldt et al. 1998 and McCoy et al. 1997). The acapulcoites tend to be fine grained (150-230 μm) and experienced low degrees of partial melting (<1%, up to a few %). The lodranites have coarser grains (540-700 μm) and experienced higher degrees of partial melting (5-10%). The ureilites are characterized by large grains averaging 1 mm in size and represent melt fractions of 10-20%. From these examples we parameterize the growth of silicate grains as a linear function from 100 μm (a typical size for small chondrules; Weisberg et al. 2006) at 0% melt fraction, increasing with a slope of 90 μm/melt %. We safely assume that grain growth is an equilibrium process: grains ripen to 5 mm (the maximum for which this model is valid, i.e. $\phi$ = 50%) in much less than $10^5$ yr (Taylor et al. 1993), faster than typical migration time scales (Eqn. 9). This parameterization is used to adjust the $\alpha$-parameter for each time step. The derivative of the melt fraction Φ in Equation 12 is calculated from Equation 6.

The total concentration of $^{26}$Al $C_{tot}$ is governed by the loss of $^{26}$Al due to melt migration and the loss due to decay in both the liquid and solid phases:

$$\frac{dC_{tot}}{dt} = -\frac{C_l}{\tau_{mig}} - \frac{C_s + C_l}{\tau_{Al}}, \qquad (13)$$

where all quantities are defined in Table 1. $C_l$ will depend on the chemical partitioning of Al into the partial melt as a function of melt fraction. It is assumed that this partitioning is an equilibrium process, because the time steps in these calculations (100s of years) are much longer than the multi-hour time scales for which Al partition coefficients are measured in the laboratory (Agee et al. 1990; Miyamoto et al. 1994; Pack et al. 2003). Chemical partitioning for equilibrium melting is described by (Best 2003):

$$C_l = \frac{C_{tot}}{D + \phi(1-D)}, \qquad (14)$$

where $D$ is the partition coefficient for Al, defined as the ratio of the weight percent of Al in the solid to the weight percent in the liquid. Irrespective of silicate composition, temperature or pressure $D$ is always much less than 1 (typical values range from 0.002 - 0.02; Agee et al. 1990; Miyamoto et al. 1994; Pack et al. 2003). We adopt a conservative value of 0.02 to minimize partitioning into the melt and thus the thermal effects of losing $^{26}$Al to melt migration.

Equation 13 can be rewritten using Equations 9, 11 and 14:

$$\frac{dC_{tot}}{dt} = -\frac{\phi^2(1-\phi)^2}{\alpha(D+\phi-\phi D)} C_{tot} - \frac{C_{tot}}{\tau_{Al}}. \qquad (15)$$

Temperature is controlled by $^{26}$Al heating and the latent heat of melting ($L$) for silicates:

$$\frac{dT}{dt} = \frac{Q_{Al}(t)}{c} - \frac{L}{c}\frac{d\Phi}{dt}, \qquad (16)$$

where $Q_{Al}(t)$ is the energy per unit mass per unit time released by the decay of $^{26}$Al and all other parameters are defined in Table 1. $Q_{Al}(t)$ can be written as:

$$Q_{Al}(t) = \frac{E_{Al}}{\tau_{Al}} C_{tot}. \qquad (17)$$

Employing the chain rule and Equation 17, we can rewrite Equation 16 as:

$$\frac{dT}{dt} = E_{Al}\left(\tau_{Al}c + L\tau_{Al}\frac{\partial \Phi}{\partial T}\right)^{-1} C_{tot}. \qquad (18)$$

The three coupled equations for $T$, $\phi$ and $C_{tot}$ (Eqns. 12, 15 and 18) are solved using the direct finite difference method (Ozisik 1994). Initial temperature $T_0$ is set to 180 K (reasonable for a planetesimal in thermal equilibrium with the solar nebula), initial melt fraction $\phi_0$ is set to 0%, and the initial $^{26}$Al concentration $C_0$ is

determined by the amount of decay before a prescribed time of instantaneous accretion:

$$C_0 = f_{Al} \left[ \frac{^{26}Al}{^{27}Al} \right] e^{-t_{acc}/\tau_{Al}}. \qquad (19)$$

Here $f_{Al}$ is the number of Al atoms per kg of unaltered planetesimal assuming an initial mass fraction of Al equal to 0.0113 (Ghosh et al. 1998), $t_{acc}$ is defined relative to the formation of CAIs and the other quantities are defined in Table 1.

### 3. Model Results

#### 3.1 Canonical Case of Conduction

Figure 1 presents a solution to Equations 3 and 4 for a body with a radius of 50 km that accreted 1.0 Myr after CAI formation. The initial and boundary temperatures were set to 180 K. Other parameters were set to their default values (Table 1). The onset of melting occurs 1.3 Myr after the formation of CAIs and at 11 Myr a peak temperature of 2615 K is reached. With the exception of the outer 4 km, this body becomes fully molten in its interior. It is shown in the next section that mobility of silicate partial melt will act to dampen internal temperatures. By 29 Myr the entire body cools to below the silicate solidus. Cooling is prolonged by the release of the heat of fusion. This canonical example is broadly consistent with the peak temperatures and timescales calculated with other thermal models (e.g. Hevey et al. 2006).

#### 3.2 The Effect of Melt Migration

Solutions from the zero-dimensional melt migration model (Sec. 2.3; Equations 12, 15 and 18) are shown in Figure 2. These simulations assume $\mu$ = 1 Pa·s, $t_{acc}$ = 1 Myr and standard values for all other quantities (Table 1). The simulations without melt migration (Fig. 2, grey curves) are identical to the temperature evolution at the center of the planetesimal in Figure 1 for $t < 6.0$ Myr, i.e. before conduction removes appreciable heat. This case results in a melt fraction of 100% and a peak temperature of 2130 K at 2.5 Myr. In the scenario with melt migration (Fig. 2, black curves) the concentration of $^{26}$Al initially decays with the half-life of the isotope; however, once partial melting begins ($T$=1373 K) Al is quickly lost from the system and subsequent heating is halted within a few times $10^5$ years. The maximum melt fraction is 27%, which results (based on our parameterization) in the growth of grains from 100 μm to 2.7 mm. This is larger by a factor of two than typical grain sizes for achondrite meteorites derived from ~10-20% melt residues (Mittlefehdt et al. 1998). We discuss possible reasons for this discrepancy in Section 4.3. The peak temperature in the simulation with melt migration is significantly lower (1610 K) than for the non-migration case.

We compute peak temperatures across a grid of viscosities and times of accretion, the two key parameters in this model of melt migration (Figure 3). If melt does not migrate, then the peak temperature depends only on the amount of energy released by the decay of $^{26}$Al, which would cause >50% melting for all times of accretion <1.7 Myr. Melt migration and removal of $^{26}$Al do not prevent the formation of >10% melt (Figure 3). However, melt migration requires that $\mu > 1$ Pa·s and $t_{acc} < 1.5$ Myr to achieve >50% melt (i.e. a magma ocean). Low viscosities (<1 Pa·s) can prevent a body from reaching melt fractions >50% (Figure 3). Such viscosities have been predicted (e.g. $\mu$=0.067 Pa·s by Folco et al. 2004) and measured in the laboratory (e.g. $\mu$<<3 Pa·s by Knoche et al. 1996) for chondritic melts.

### 3.3 Thermal and Petrologic Evolution of the Crust

Migration of Al-enriched melts to the surface would produce an $^{26}$Al-enriched crust. If partial melting and crust formation occur early, i.e. within one or two $^{26}$Al half-lives, this concentration could re-melt the crust. Analogous concentration of long-lived radionuclides following differentiation has been considered as a possible contributor of pre-basin and/or basin lunar mare volcanism (Manga et al. 1991; Arkani-Hamed et al. 2001) and relatively recent volcanism on Mars (Schumacher et al. 2007). Assuming a planetesimal average melt fraction $F$, and completely efficient crust formation, the crust thickness $d$ will be $RF/3$. To conservatively estimate the condition for re-melting, we assume that the equilibrium surface temperature of the planetesimal is much less than the silicate melting point, and that the eruption happened slowly enough for the crust to initially and everywhere reach that equilibrium temperature. In the plane-parallel approximation, the thickness of the crust's thermal boundary layer at any time is (Turcotte et al. 2002):

$$\delta \sim 2(\kappa t)^{1/2}, \quad (20)$$

and is independent of surface temperature or heating rate. Penetration of the cold thermal boundary layer to the center of the crust (i.e. $\delta = d/2$) will occur after an elapsed time:

$$\Delta t \approx \frac{1}{\kappa}\left(\frac{RF}{12}\right)^2. \quad (21)$$

The temperature elevation in the center of the crust will reach a maximum at about this time and is approximately

$$\Delta T_{peak} \approx \frac{Q_0}{c_p}\int_{t_C}^{t_C+\Delta t}e^{-t'/\tau_{Al}}dt' = \frac{Q_0 \tau_{Al}}{c_p}\left[1-\exp\left(-\frac{1}{\kappa \tau_{Al}}\left(\frac{RF}{12}\right)^2\right)\right]\exp\left(-\frac{t_C}{\tau_{Al}}\right), \quad (22)$$

where $t_C$ is the formation time of the crust relative to CAIs. For standard values of the parameters, re-melting of a crust that formed soon (<< 1 Myr) after CAIs would occur on planetesimals where *RF* exceeded 20 km (e.g. 200 km radius for a 10% average melt fraction). No re-melting would occur on any planetesimal (i.e. the required *RF* diverges) with a crust that formed later than 2 Myr after time zero. Re-melting of basalt would produce relatively Si-rich, Na-rich and Mg-poor rocks resembling andesitic basalts or trjondhhemites (Helz 1976; Martin 1986; Petford et al. 1996).

*3.4 Differentiation following the loss of $^{26}$Al*

Differentiation into a mantle and Fe-Ni core (like those represented by the iron meteorites) requires at least 50% silicate partial melting (Taylor 1992), which corresponds to a temperature of 1558 K (Eqn. 6). Here we estimate whether the decay of $^{60}$Fe was sufficient to reach this temperature following silicate melting and the loss of $^{26}$Al via melt migration (Sec. 3.3). We use the conduction model (Sec. 2.1) to compute the thermal evolution of planetesimals 100 km in diameter, a typical size for the parent bodies of iron meteorites (Haack et al. 2004). We assume that once temperatures reach the silicate solidus, all of the $^{26}$Al is instantaneously removed by migration, i.e. fast (~$10^5$ years, Fig. 2) relative to the mean life of $^{60}$Fe. Pyroclastic eruptions at the surface may have facilitated the complete removal of basaltic melt from a planetesimal (Wilson et al. 1991). The initial abundance and decay properties of $^{60}$Fe are given in Table 1.

Figure 4 shows the thermal evolution of four different planetesimals with formation times ranging from 0 to 2 Myr. In all cases, the silicate solidus is reached within a few times $10^5$ years after the start of each simulation. The melt fraction does not reach 50% for formation times greater than ~2 Myr, consistent with Hf-W ages for the formation of magmatic iron meteorites (Goldstein et al. 2009). Core formation in the presence of melt fractions >50% could take place in as little as a few years (Stevenson 1990; Taylor 1992), thus the persistence of high temperatures over Myr timescales (Fig. 4) suggests that differentiation was likely for 100 km bodies when $t_{acc}$ < 2 Myr.

*3.5 Lower Size Limit of Differentiated Bodies*

The time scale for the formation of a basaltic crust is independent of size (Sec. 2.2), provided temperatures remain high enough to sustain partial melting. Thus the minimum size for differentiation will be set by the rate of conductive heat loss. The characteristic time scale for thermal conduction into a semi-infinite half space is $R^2/\kappa$ (Turcotte et al. 2002); however, this is a poor approximation for internally heated spheres, which cool several times faster (Osuga 2000). Furthermore, the correlation between effective heat capacity and melt fraction (Sec. 2.1) means that $R^2/\kappa$ is not constant for a body that reaches melting temperatures. Thus, characteristic cooling times for planetesimals reflect a mix of conductive heat loss, the decaying $^{26}$Al and $^{60}$Fe heat sources, variable heat capacity, and the influence of the heat of fusion on the thermal budget.

E-folding times (defined as $\tau_{cool}$, the time it takes the peak temperature at the center of a planetesimal to drop by a factor of 1/e) are numerically computed with the conduction model (Sec. 2.1) for bodies with radii between 1 and 250 km. Internal heat production is maximized by setting $t_{acc}$ = 0 Myr. These calculations show that characteristic cooling times for planetesimals with radii >10 km are a factor of ~3 less than $R^2/\kappa$ when $\kappa$ is fixed at 7.9×10$^{-7}$ m$^2$/s (i.e. $c$ = 800 J/kg/K, $\rho$ = 3300 kg/m$^3$ and $k$ = 2.1 W/m/K; Eqn. 2). Smaller bodies ($R$ <5 km) require longer than $R^2/\kappa$ to cool due to the internal heat sources. However, the peak temperatures of these small bodies do not become large: a planetesimal with a radius of 1 km and $t_{acc}$ = 0 Myr will reach a peak temperature of 245 K, only 65 K above the ambient and starting temperature, but then requires 1.1 Myr to cool by a factor of 1/e. Smaller melt fractions from later times of accretion reduce the effective heat capacity and thus decrease $\tau_{cool}$.

The minimum size for differentiation will occur when the heat flux is maximum, i.e. $t_{acc}$ = 0 Myr. The approximate conductive cooling time for a planetesimal with $t_{acc}$ = 0 Myr is given by a least-squares fit to our numerically computed e-folding times:

$$\tau_{cool}(t_{acc}=0) \approx 0.014\, Myr \left(\frac{R}{1 km}\right)^2. \qquad (23)$$

Planetesimals with $\tau_{cool}$ longer than the most important heating time scale $\tau_{Al}$ will sustain melting temperatures and differentiate. Equation 23 is equal to $\tau_{Al}$ for bodies 18 km in size.

## 4. Neglected and/or Unimportant Processes

### *4.1 Convection*

Our models assume that the only significant subsolidus mechanism of heat transport is conduction. Subsolidus convection in a uniformly heated body will occur only if a thermal boundary layer becomes unstable to convection. While the interior of the planetesimal heats up, the surface will remain cool and a conducting boundary layer of thickness $\delta$ (Eqn. 20) will appear and thicken. The cooler, negatively-buoyant boundary layer may eventually become Rayleigh-Taylor unstable, and when the viscosity contrast across the boundary layer is large (i.e. due to temperature dependence), the instability manifests itself with short-wavelength "drips" that grow from the lower surface, break off, and descend into the warmer, less dense interior (Jaupart et al. 1985). This will drive a passive upwards counter-flow and maintain a *subadiabatic* temperature gradient below the boundary layer (Moore 2008). Due to erosion at its base, the boundary layer ceases to grow and the heat flow will be set by conduction through its equilibrium thickness.

In the plane-parallel and thin boundary layer ($\delta<R$) approximations, the temperature profile with depth $z$ in a body is expressed as an integral of Green's function solutions to the thermal diffusion equation:

$$T = T_s + \frac{Qt}{c}\int_0^1 d\tau \cdot erf\left(\frac{\zeta}{\sqrt{1-\tau}}\right), \tag{24}$$

where $\zeta = z/(\kappa t)^{1/2}$, $\tau$ is dimensionless time, and all other parameters are given in Table 1. The criterion for the onset of the Rayleigh-Taylor instability is usually described as a local Rayleigh number $Ra$ exceeding a critical value $Ra_c$, where

$$Ra = \frac{\alpha \rho g \Delta T \delta^3}{\kappa \mu_B}, \tag{25}$$

with $\Delta T$ the temperature drop across the boundary layer. In the case of a material with a strongly temperature-dependent viscosity, only the lower, warmer, and less viscous part of the boundary layer will participate in the instability. This complexity can be accommodated by calculating an "available buoyancy" (Conrad et al. 1999) but this correction is small (Korenaga et al. 2003) compared to large uncertainties in the viscosity itself and we ignore this complication and instead compute a maximum $Ra$ that favors the onset of solid-state convection. Substituting Equation 20, $\Delta T = Qt/c$, and surface gravity for a homogeneous body into Equation 25, we obtain a critical time $\tau_c$ for the formation of convective instabilities:

$$\tau_c = \left[\frac{3Ra_c c \mu_B}{32\pi\rho^2 \alpha GRQ\sqrt{\kappa}}\right]^{2/5}. \tag{26}$$

$Ra_c$ lies between 1000 and 2000 (Korenaga et al. 2003). The parameter here with the largest uncertainty is $\mu_B$, the viscosity at the base of the lithosphere, which will depend on temperature, and fall markedly as the planetesimal heats up and begins to melt. The viscosity of Earth's deep mantle is $\sim 10^{22}$ Pa·s but beneath mid-ocean ridges where partial melting is occurring it may be as low as a few times $10^{17}$ to $10^{19}$ Pa·s (Pollitz et al. 1998; James et al. 2008). In the most optimistic scenario ($\mu_B = 10^{17}$ Pa·s, $t_{acc} = 0$ Myr and $R = 1000$ km) $\tau_c = 0.27$ Myr. From the thermal conduction model (Sec. 2.1), a 1000 km body with $t_{acc} = 0$ Myr reaches silicate-melting temperatures in only 0.13 Myr, well before the onset of subsolidus convection. In fact, for all of the scenarios considered in Section 3 (i.e. $t_{acc} < 2.0$ and R ≤ 50 km), $\tau_c$ is longer than the time to achieve silicate melting. Only large (>100 km), late accreting (>2 Myr) bodies will be affected by subsolidus convection, thus justifying our use of a conduction-only model.

Vigorous convection would occur at temperatures well above the solidus, i.e. in a magma ocean (Taylor et al. 1993). However, our focus on constraining the conditions leading up to differentiation does not require treatment of magma ocean processes. It is sufficient to assume that differentiation will occur (either by fractional or equilibrium crystallization) in the presence of high degrees of melting.

*4.2 Sintering*

The conduction model (Sec. 2.1) does not include the effects of sintering of silicate grains, expected to occur around 700 K and to produce a rapid increase in ρ and a corresponding jump in thermal conductivity from initial values as low as 0.001 W/m/K (Hevey et al. 2006; Sahijpal et al. 2007). Computing peak central temperatures with a modified conduction model demonstrates that sintering does not have a significant effect on our results. The peak temperature reached at the center of a 50 km body with $t_{acc}$ = 2.2 Myr and fixed $k$ = 2.1 W/m/K is 1480 K. However, a body originally 63 km in radius with $k$ = 0.001 W/m/K that compacts down to 50 km (i.e. the loss of 50% pore space) and $k$ = 2.1 W/m/K at 700 K will reach a nearly identical peak temperature of 1484 K. This treatment of sintering is achieved by allowing κ (Eqn. 2) to vary with temperature. The primary effect of sintering is that the cool outer layers remain uncompacted and thus decrease the rate of interior cooling. It is the time of accretion (i.e. the initial $^{26}$Al and $^{60}$Fe abundances) that largely controls peak temperature.

Although this treatment of sintering shows little effect on bodies that reach melting temperatures (>>700 K), there are large uncertainties associated with the assumed initial porosities (and the corresponding thermal conductivities). A value of $k$ = 0.001 W/m/K is based on laboratory measurements of the lunar regolith in a vacuum (Fountain et al. 1970). However, processes such as volatile release and collisional packing during accretion would act to decrease vacuum space and thus increase conductivity. Other differences, such as size, collisional history and composition, suggest that the lunar regolith may not be an ideal analog for an unsintered planetesimal.

*4.3 Changes in Planetesimal Hydraulics*

The following phenomena would affect the mobility of melts in the melt migration model (Sec. 2.3): an increase in melt bouyancy due to exsolution of volatiles (Wilson et al. 1991; Keil et al. 1993); an increase in the effective grain size and decrease in the resistance to flow due to coalescence of melt channels into larger veins and dikes (Wilson et al. 2008); and fracturing due to volume changes associated with serpentinization reactions (Cohen et al. 2000), sintering (Hevey et al. 2006) and the production of partial melt. In general, these phenomena would increase permeability, thus Equation 9 represents a conservative estimate for the rate of melt migration. An increase in the efficiency of drainage of Al-enriched melt would produce less melting and (due to our parameterization of grain size as a function of melt fraction) a smaller final grain size. This is supported by meteoritic evidence, which suggests that efficient draining of all liquid occurs once grain sizes become ~1 mm and melt fractions exceed 20% (McCoy et al. 1997).

### 4.4 Core Formation

This work focuses on the behavior of silicate melts and does not treat the formation of Fe-Ni alloy cores. Theoretical studies suggest that core formation would proceed quickly (much faster than the mean life of $^{26}$Al), but required melt fractions >40% (Stevenson 1990; Taylor 1992). Laboratory experiments produce conflicting results regarding the efficiency of metal-silicate segregation at lower melt fractions (see McSween et al. 1978; Rushmer et al. 2000; Yoshino et al. 2003; 2004; McCoy et al. 2006 and references within for details). In short, a complicated interplay of temperature, pressure, alloy melt composition, silicate melt fraction, and rheology of the matrix dictated the way in which cores formed, e.g. percolation of liquid metal through interconnected pore spaces between silicate grains versus gravitational settling of metal globules through a viscous silicate matrix (Stevenson 1990).

Core formation would have concentrated $^{60}$Fe and $^{26}$Al into smaller volumes (the core and mantle respectively). Chondritic materials typically have ~20 weight % Fe (Lodders et al. 1998), so that the volume of an Fe-Ni core would occupy less than 1/8 the total volume of a differentiated planetesimal. Thus, core formation would result in an enhanced number density per kg of $^{26}$Al in the mantle of no more than 15% over chondritic concentrations. This enhancement is small relative to other uncertainties in our model (e.g. grain size distribution, composition, accretion scenario) and is probably insignificant. The concentration of $^{60}$Fe into a core can increase cooling times due to the insulating effects of the surrounding silicate matrix; however, the overall thermal budget of a planetesimal will be unchanged. The gravitational potential released by core formation is insignificant for planetesimal-size bodies (Moskovitz 2009).

### 4.5 Temperature-Dependent Heat Capacity

We assume the specific heat capacities of solid and liquid phases are independent of temperature and are equal to 800 and 2000 J/kg/K respectively. Although heat capacity generally increases with temperature, these values are near the upper limits for chondritic assemblages (Ghosh et al. 1999; Navrotsky 1995) and thus result in conservative estimates for peak temperatures. Inserting lower limits of 600 and 1700 J/kg/K (Navrotsky 1995) for a 100 km body with $t_{acc}$ = 1.0 Myr results in a peak temperature of 2850 K, an increase of ~240 K over the standard scenario presented in Figure 1. This illustrates that the assumed heat capacities produce temperatures that are underestimated by no more than 10%.

### 4.6 Non-instantaneous accretion

All simulations presented here assume instantaneous accretion. Merk et al. (2002) showed that a linear accretion rate of 200 km/Myr significantly affected peak temperatures. Ghosh et al. (2003) showed that average accretion rates of 15-90 km/Myr affected heating and cooling timescales. However, a recent U-Pb age for the

IVA iron meteorites (Blichert-Toft et al. 2010) suggests that their parent body accreted, differentiated and cooled to Pb isotopic closure (~600 K) in less than 3 Myr. For any body with R>15 km, which applies to the IVAs and in fact all iron meteorite parent bodies (Haack et al. 2004), accretion had to occur in much less than $10^6$ years to allow sufficient time for cooling (Eqn. 23). Recent dynamical studies that treat the turbulent concentration of small particles in proto-planetary disks (Johansen et al. 2007; 2009; Cuzzi et al. 2010) show that planetesimals can grow to sizes of 100 km or larger on orbital timescales (typically < 100 years). Such rapid accretion is nearly instantaneous relative to other relevant timescales (e.g. $\tau_{mig}$, $\tau_{cool}$, $\tau_{Al}$) and thus does not affect thermal evolution scenarios.

If we had considered some parameterization of prolonged accretion, our standard conduction model (Fig. 1) would likely reach peak temperatures lower by a few hundred K (Merk et al. 2002; Ghosh et al. 2003). Slower accretion would require even earlier formation times for a body to melt and differentiate, particularly in the absence of $^{26}$Al (Fig. 4). The consequences of melt migration (Sec. 3.2) are independent of accretion scenario as long as melting temperatures are reached. It is possible that melt migration would occur before a planetesimal had fully formed, particularly for accretion times longer than $\tau_{mig}$. For a given initial planetesimal melt fraction, the re-melting of a crust only depends on its time of formation relative to CAIs (Sec. 3.3) and thus is independent of the details of accretion. The minimum size for differentiation (Sec. 3.5) is unaffected by accretion scenario, because the limiting case occurs with instantaneous accretion at $t_{acc}$ = 0.

## 5. Synthesis and Discussion

We have presented two models for the thermal evolution of planetesimals. The first is a spherically symmetric conduction model that account for SLR decay and the thermodynamics of phase transitions (Sec. 2.1). The second is based on a parameterization for the rate of buoyant melt migration (Sec. 2.2) and calculates changes in the rate of heating due to transport and removal of $^{26}$Al in partial melts (Sec. 2.3). This migration model was used to show that low viscosity ($\mu$ < 1 Pa·s) silicate melt can migrate on timescales less than the mean life of $^{26}$Al (Sec. 3.2). Following such migration, the residual material becomes enriched in coarse-grained silicates and Fe-Ni metal, depleted in elemental Al, and depleted in low temperature phases like plagioclase. The lodranite and ureilite achondrites are examples of such melt residues (Mittlefehdt et al. 1998). After removal of $^{26}$Al from the interior of a planetesimal, subsequent heating from the decay of $^{60}$Fe alone was sufficient to generate melt fractions >50% and complete differentiation as long as accretion occurred within 2 Myr of CAI formation (Sec. 3.4). This is in contrast to previous studies that either adopted lower $^{60}$Fe abundances (e.g. Ghosh et al. 1998), leading to the conclusion that $^{60}$Fe did not have an appreciable effect on thermal evolution, or did not consider heating by $^{60}$Fe following the removal of $^{26}$Al (e.g. Hevey et al. 2006). We showed that remelting of a basaltic crust enriched in $^{26}$Al would have occurred for bodies that accreted within 2 Myr of CAIs, reached melt fractions of order 10%, and were ≳100 km in size (Sec. 3.3). Finally, the conduction model was used to calculate a lower size limit for differentiation of 18 km (Sec. 3.5).

We distill our calculations of planetesimal heating, melting, and differentiation into Figure 5. The maximum radius in this figure (250 km) is that of asteroid 4 Vesta. The one-dimensional conduction model (Sec. 2.1) was run for every location in this $R$-$t_{acc}$ parameter space. The maximum temperatures at depths of 50% of the radii were recorded for every $R$-$t_{acc}$ pair. The three black, solid curves in this figure represent objects whose peak temperatures at 50% radius were equal to the Fe-S eutectic (1200 K), the silicate solidus (1400 K) and the silicate liquidus (1900 K). These curves segregate this figure into two primary domains. To the right of these curves, no melting would occur. To the left, high degrees of melting would result in surface volcanism and a vigorously convecting magma ocean (Taylor et al. 1993, Righter et al. 1997). Our model predicts that silicate melting required accretion within 2.7 Myr of CAIs (Fig. 5), consistent with radiometric dating of the magmatic iron meteorites, whose parent bodies accreted and differentiated within 1-2 Myr of CAI formation (Markowski et al. 2006). This constraint is slightly less than the 2.85 Myr for 25% melting of the Vesta parent body calculated by Ghosh et al. (1998). Their higher initial temperature (by more than 100 K) is the likely cause of this discrepancy.

The shaded regions in Figure 5 represent different fates of molten material. The light grey region represents planetesimals too small (<20 km) for compaction-driven melt migration to occur, i.e. $\delta_c$ is comparable to or larger than $R$. These bodies would not differentiate by melt migration, but could form a magma ocean. This region can extend to larger radii (perhaps up to $R$=50 km) depending on the specific grain size distribution and rheologic properties of the planetesimal (Sec. 2.2). The medium grey region represents bodies small enough (<100 km) that pyroclastically erupted melt can escape (Wilson et al. 1991; Keil et al. 1993). This ejection of melt would have precluded the formation of a basaltic crust and may be relevant to understanding the scarcity of basaltic material amongst asteroids and meteorites in the present day (Moskovitz et al. 2008; Burbine et al. 1996). The dark grey region represents larger objects, e.g. Vesta, where melt does not escape and instead is deposited in a crust. Rapid removal of $^{26}$Al-bearing melt from the interior of planetesimals in the medium and dark grey regions will significantly alter their peak temperatures (Sec. 3.2), though the presence of $^{60}$Fe may have helped these bodies to fully differentiate (Sec. 3.4). The arrow at top indicates the upper limit to the time of accretion (2 Myr) for which a planetesimal depleted in $^{26}$Al could still have differentiated as a result of the decay of $^{60}$Fe.

Subsolidus convection did not occur for bodies smaller than ~100 km, thus justifying the use of a conduction model. However, later times of accretion decrease the critical radius at which subsolidus convective instabilities developed prior to melting. The hatched region in Figure 5 represents this limit. The lower boundary of this region is defined by the $R$-$t_{acc}$ pairs for which $\tau_c$ (Eqn. 26) is equal to the time required to reach melting, assuming a bulk viscosity of $10^{17}$ Pa·s, a reasonable lower limit to the viscosity of the Earth's asthenosphere (Pollitz et al. 1998; James et al. 2008). Convective overturn on these bodies would act to reduce peak temperatures and prevent melting. The bulk viscosity of material near partial melting temperatures can vary greatly. If bulk viscosities were as low as $\sim 10^{16}$ Pa·s then convective instabilities would affect over half of the melting region in Figure 5. Though Hevey et al. (2006) and Sahijpal et al. (2007) parameterized fluid convection in a magma ocean, we are unaware of any study that has considered the thermal consequences of subsolidus convection on planetesimal-size bodies.

The dashed curve in Figure 5 indicates the $R$-$t_{acc}$ pairs for which the melt migration time scale ($\tau_{mig}$) and the conductive cooling time scale ($\tau_{cool}$) are equal, assuming a grain size of 1 mm, melt viscosity of 1 Pa·s and melt fraction of 10%. This boundary is sensitive to variations in these quantities and can shift across the full range of radii. Above the dashed line, removal of Al-enriched melt is the dominant cooling process. Below, the conductive loss of heat causes melt (if any is present) to freeze in place before migration occurs. Irrespective of the dominant cooling mechanism, if $\tau_{cool} < \tau_{Al}$ then a planetesimal will not melt.

Our calculations for the rate of compaction-driven melt migration (Sec. 2.2) and the efficiency of conductive cooling (Sec. 3.5) suggest that differentiation only occurred for bodies larger than $\sim 20$ km in diameter, consistent with previous studies (e.g. Hevey et al. 2006; Sahijpal et al. 2007). In the present day Solar System, the only asteroids other than 4 Vesta that are >15 km in diameter (Delbo et al. 2006; Tedesco et al. 2002) and have spectroscopically confirmed basaltic surfaces are 1459 Magnya and 1904 Massevitch. Although these bodies are near the differentiation limit, it is not clear whether they could escape collisional disruption over the age of the Solar System (Bottke et al. 2005).

In general, our results are most consistent with Sahijpal et al. (2007), though several key differences do exist. Most notably is their treatment of melt migration velocities. They prescribe migration rates of 30-300 m/yr, however these rates are not physically motivated and instead are a reflection of stability requirements in their numerical model. These rates are more than an order of magnitude faster than our most optimistic scenario (i.e. 1 cm grains with 10's of % partial melt) and thus overestimate the rate of $^{26}$Al redistribution. Another difference with Sahijpal et al. (2007) is that we do not include changes to volume and thermal conductivity associated with sintering at 700 K (Hevey et al. 2006). Implementation of our conduction model with a temperature-dependent κ shows that the high peak temperatures (>700 K) in which we are interested are barely affected by sintering (Sec. 4.3). The most significant effect of sintering for bodies of all sizes is an increase of interior cooling rates due to the insulating effects of an unsintered, low conductivity surface layer.

Our results raise several issues in need of further investigation. For instance, the basaltic eucrite meteorites are thought to have formed from liquid expelled out of a partial melt region (Consolmagno et al. 1977; Walker et al. 1978, Righter et al. 1997). This scenario probably required mm-size grains and melt viscosities ~1 Pa·s (Sec. 3.2). Experimental measurement of the viscosity of basalts with compositions similar to eucrites can test the plausibility of this model. Calculations of the chemical composition of chondritic melts as a function of temperature (e.g. using the MELTS software package, Ghiorso et al. 1995) could also constrain their viscosity evolution and thus their resistance to migration. Lastly, our calculations suggest that bodies as small as ~12 km in diameter could partially melt (Fig. 5). We do not expect melts to segregate on these bodies due to their small size (Sec. 2.2). However, fractional crystallization in a magma ocean could produce a differentiated structure. Additional modeling of magma ocean formation and evolution on bodies with low self-gravity would address whether small basaltic asteroids in the present day Solar System, e.g. 1459 Magnya and 1904 Massevitch, are intact differentiated planetesimal or relic fragments from heavily eroded parent bodies.

*Acknowledgments* We are grateful to G. J. Taylor, T. McCoy and J. Sinton for their insightful opinions regarding this manuscript and appreciate the opportunity to have benefitted from their vast expertise in this subject matter. We thank H. McSween and L. Wilson for their helpful reviews. N.M. would like to acknowledge support from the Carnegie Institution of Washington, the NASA Astrobiology Institute and NASA GSRP grant NNX06AI30H.

1197    **Table 1** Symbols and definitions.

| Symbol | Name, definition and reference | Value [Units] |
|---|---|---|
| $T$ | Temperature | [K] |
| $T_s$ | Surface temperature | [K] |
| $t$ | Time | [s] |
| $t_{acc}$ | Time of accretion | [s] |
| $\phi$ | Net melt fraction by weight | - |
| $\Phi$ | Instantaneous amount of melt | - |
| $R$ | Radius of body | [m] |
| $r$ | Radial variable | [m] |
| $\delta$ | Thermal boundary layer thickness, $= 2(\kappa t)^{1/2}$ | [m] |
| $a$ | Grain size | [m] |
| $\mu$ | Liquid viscosity | [Pa·s] |
| $\mu_B$ | Effective bulk viscosity | [Pa·s] |
| $\Delta\rho$ | Density contrast between solid and liquid | 300 kg/m$^3$ |
| $\overline{\rho}$ | Mean density of olivine | 3300 kg/m$^3$ |
| $Q(t)$ | SLR decay energy generation rate (Eqn. 5) | [J/kg/s] |
| $f_{Al}$ | Chondritic abundance of Al (Ghosh et al. 1998) | $2.62 \times 10^{23}$ kg$^{-1}$ |
| $E_{Al}$ | $^{26}$Al decay energy per atom (Lide 2002) | $6.4154 \times 10^{-13}$ J |
| [$^{26}$Al/$^{27}$Al] | $^{26}$Al initial abundance (Jacobsen et al. 2008) | $5 \times 10^{-5}$ |
| $\tau_{1/2}$ | $^{26}$Al half-life (Goswami et al. 2005) | 0.74 Myr |
| $\tau_{Al}$ | $^{26}$Al mean life, $= \tau_{1/2} / \ln(2)$ | 1.07 Myr |
| $f_{Fe}$ | Chondritic abundance of Fe (Ghosh et al. 1998) | $2.41 \times 10^{24}$ kg$^{-1}$ |
| $E_{Fe}$ | $^{60}$Fe decay energy per atom (Ghosh et al. 1998) | $4.87 \times 10^{-13}$ J |
| [$^{60}$Fe/$^{56}$Fe] | $^{60}$Fe initial abundance (Tachibana et al. 2006; Dauphas et al. 2008; Rugel et al. 2009; Mishra et al. 2010) | $6 \times 10^{-7}$ |
| $\tau_{Fe}$ | $^{60}$Fe mean life (Rugel et al. 2009) | 3.49 Myr |
| $C_l$ | Concentration of $^{26}$Al in the liquid phase | [1/kg] |
| $C_s$ | Concentration of $^{26}$Al in the solid phase | [1/kg] |
| $C_{tot}$ | Total concentration of $^{26}$Al, $= C_l + C_s$ | [1/kg] |
| $D$ | Al partition coefficient | - |
| $\alpha$ | Thermal expansion coefficient | [1/K] |
| $c$ | Specific heat at constant pressure | [J/kg/K] |
| $k$ | Thermal conductivity | [W/m/K] |
| $\kappa$ | Thermal diffusivity, (Eqn. 2) | [m$^2$/s] |
| $L$ | Latent heat of fusion for silicates (Ghosh et al. 1998) | 400 kJ/kg |
| $Ra$ | Raleigh number (Eqn. 25) | - |
| $\tau_c$ | Convective instability timescale (Eqn. 26) | [s] |
| $\tau_{mig}$ | Melt migration time scale (Eqn. 9) | [s] |
| $\tau_{cool}$ | Conductive cooling timescale (Sec. 3.5) | [s] |
| $g$ | Gravitational acceleration, $= (4\pi G \overline{\rho})/3$ | [m/s$^2$] |
| $G$ | Gravitational constant | $6.673 \times 10^{-11}$ m$^3$/kg/s$^2$ |

Moskovitz
Figure 1

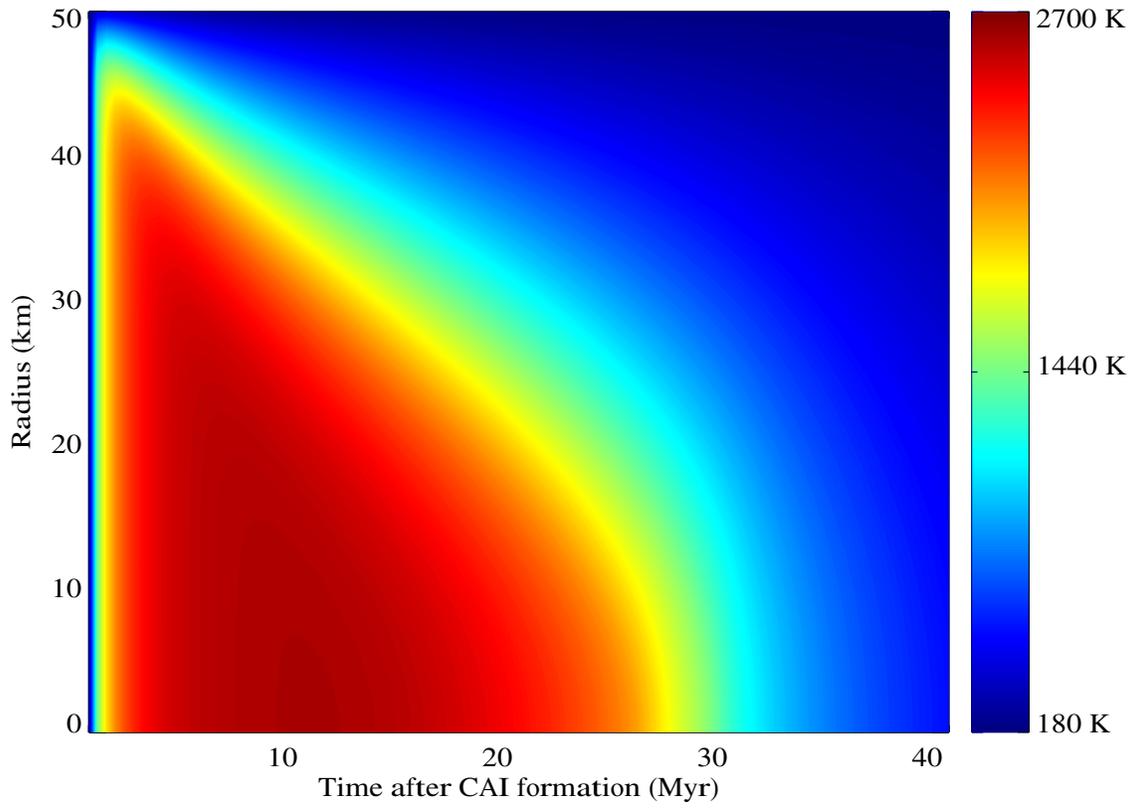

Moskovitz Figure 2

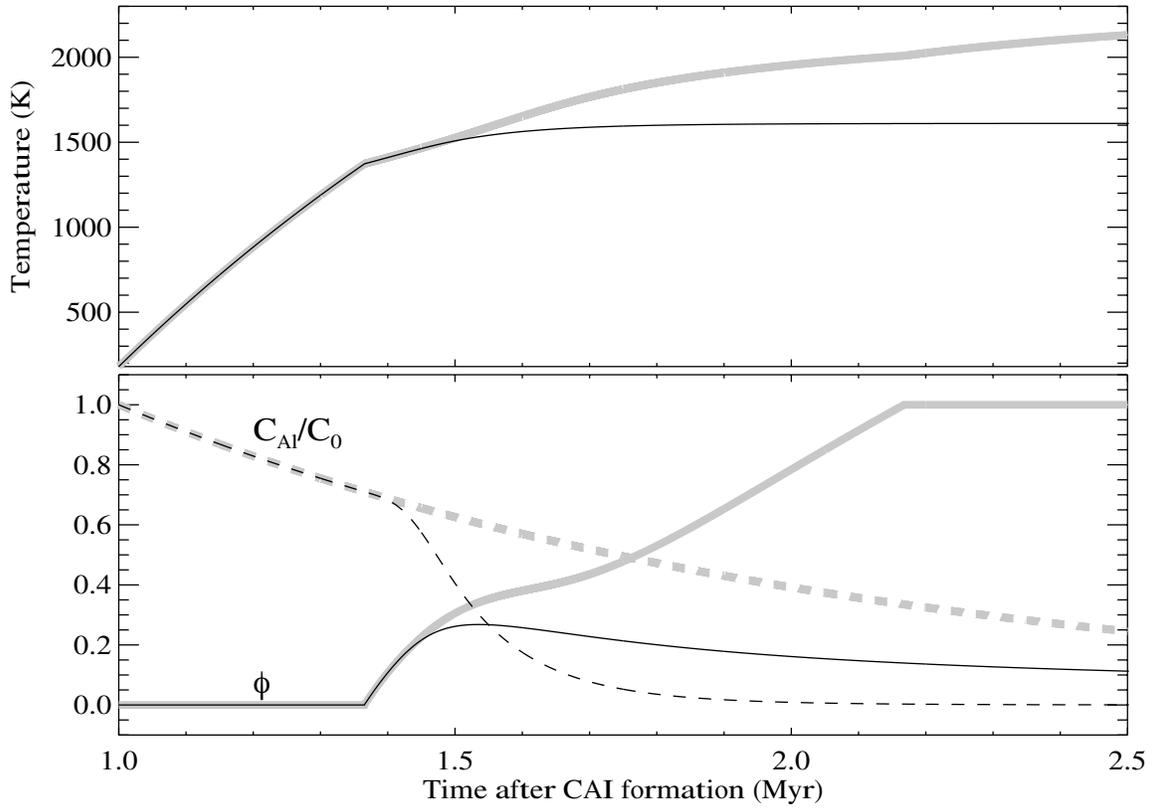

1200

Moskovitz Figure 3

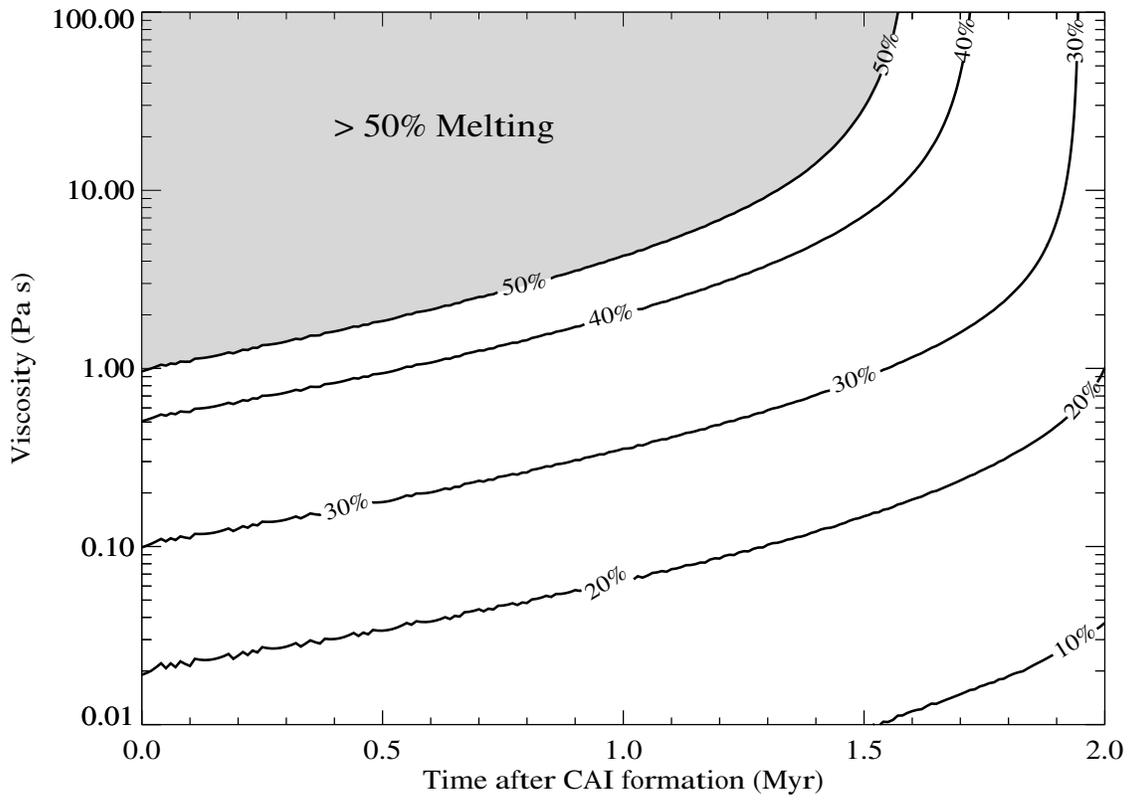

1201

Moskovitz
Figure 4

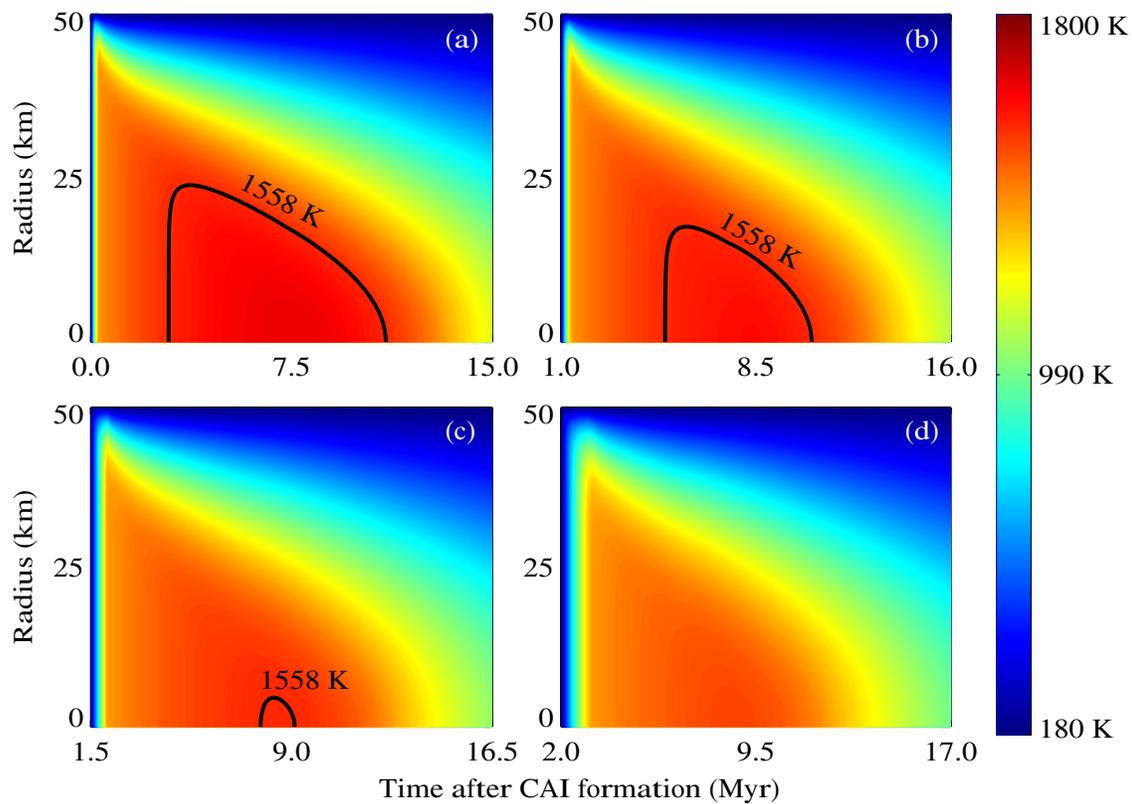

1202
1203

Moskovitz
Figure 5

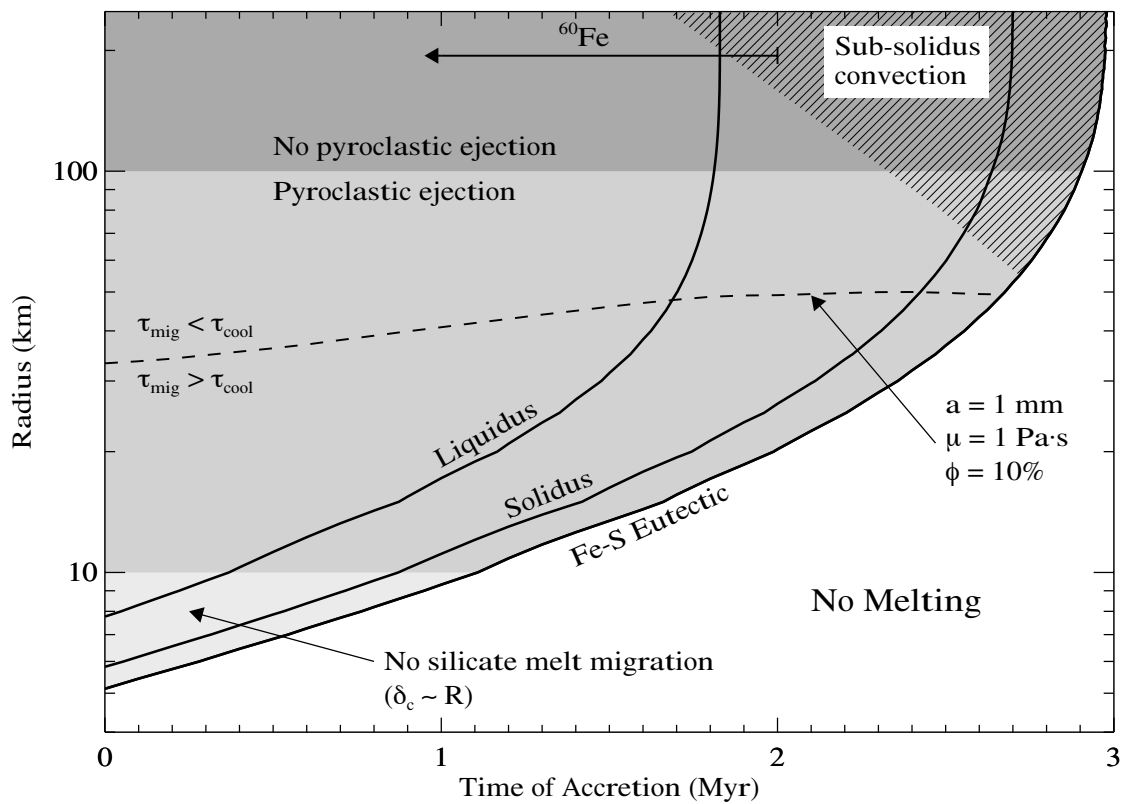

1204
1205

**Figure Captions**

Fig. 1. Temperature evolution for an $R$ = 50 km planetesimal that accreted 1.0 Myr after CAI formation. Time is reported relative to CAI formation. The body reaches its maximum temperature 10 Myr after accretion and then cools on a characteristic time scale ($\tau_{cool}$) of ~21 Myr.

Fig. 2. Melt migration simulation for $t_{acc}$ = 1 Myr. The top panel shows the temperature evolution, the bottom shows the melt fraction ($\phi$, solid) and the normalized $^{26}$Al concentration ($C_{Al}/C_0$, dashed). The grey curves assume no melt migration and are identical to the temperature evolution at the center of the body in Figure 1 for $t$ < 2.5 Myr. The black curves include melt migration. At the onset of melting (T=1373 K) the two scenarios diverge. Without migration the melt fraction reaches 100% and the peak temperature is 2130 K at 2.5 Myr. With migration the melt fraction is never larger than 27% and the peak temperature is only 1610 K. The changes of inflection in these curves are due to the latent heat of melting and the parameterization of melt fraction (Eqn. 6).

Fig. 3. Melt fraction as a function of viscosity and time of accretion. The contours represent 10% intervals of partial melting (1407, 1452, 1522, 1707, and 1830 K; McKenzie et al. 1988). Migration of low viscosity melt acts to remove $^{26}$Al and reduce peak temperature. Without melt migration, all times of accretion <1.7 Myr would result in >50% melting. Combinations of viscosity and accretion time that lie above the 50% melting line would produce conditions favorable to the formation of a magma ocean. The jagged features in the melt contours are numerical artifacts.

Fig. 4. Temperature evolution due to $^{60}$Fe decay in Al-depleted planetesimals. The four panels correspond to different times of instantaneous accretion relative to CAI formation: 0, 1.0, 1.5, and 2.0 Myr (a-d respectively). In all cases, heating by the decay of $^{26}$Al is stopped as soon as the silicate solidus (1373 K) is reached, which typically occurs within a few times $10^5$ years. If melt migration removed $^{26}$Al from the parent bodies of the magmatic iron meteorites, then they had to have accreted within 2 Myr of CAI formation in order to fully differentiate (i.e. reach 50% partial melting at 1558 K).

Fig. 5. Thermal evolution as a function of planetesimal radius ($R$) and time of accretion ($t_{acc}$, relative to CAI formation). The solid curves represent contours of maximum temperature at 50% radius (liquidus = 1900 K, solidus = 1400 K, Fe-S eutectic = 1200 K). The arrow at top indicates an upper limit to the time of accretion for $^{60}$Fe-driven differentiation to occur. The light grey region represents bodies for which compaction-driven melt migration will not occur because the size of the compaction region ($\delta_c$) is comparable to $R$. Bodies in the medium grey region can lose melt via pyroclastic eruption, those in the dark grey region have sufficient self-gravity to retain melt erupted at the surface (Wilson et al. 1991; Keil et al. 1993). The dashed line represents bodies for which the melt migration time scale ($\tau_{mig}$) is equal to the conductive cooling time scale ($\tau_{cool}$), assuming melt parameters $a$=1 mm, $\mu$=1 Pa·s and $\phi$=10%. The hatched region in the upper right represents bodies that experience solid-state convection if their subsolidus viscosity is > $10^{17}$ Pa·s.